\documentclass[aps,pro,superscriptaddress,twocolumn,floatfix]{revtex4}
\usepackage{amsfonts,color}
\usepackage{amsmath}
\usepackage{amssymb}
\usepackage{graphicx}
\usepackage{dsfont}

\begin{document}

\title{Integrable model of a $p$-wave bosonic superfluid}

\author{Sergio Lerma-Hern\'andez}
\affiliation{Facultad de F\'{i}sica, Universidad Veracruzana, Circuito Aguirre Beltr\'an s/n, Xalapa, Veracruz 91000, Mexico}

\author{Jorge Dukelsky}
\affiliation{Instituto de Estructura de la Materia, CSIC, Serrano 123, 28006 Madrid, Spain}

\author{Gerardo Ortiz}
\affiliation{Department of Physics, Indiana University, Bloomington, Indiana 47405, USA}

\begin{abstract}
We present an exactly-solvable $p$-wave pairing model for two bosonic
species. The model is solvable in any spatial dimension and shares some
commonalities with the $p + ip$ Richardson-Gaudin fermionic model, such
as  a third order quantum phase transition. However, contrary to the
fermionic case,  in the bosonic model  the transition separates a
gapless fragmented singlet pair condensate from a pair Bose superfluid,
and the exact eigenstate at the quantum critical point is a pair condensate
analogous to the fermionic Moore-Read state.
\end{abstract}

\pacs{74.90.+n, 74.45.+c, 03.65.Vf, 74.50.+r}
\maketitle

\section{Introduction}

Integrable Richardson-Gaudin (RG) models \cite{Amico2001, Dukelsky2001}
based on the $su$(2) fermion pair algebra have attracted a lot of attention in recent years.
Starting with studies of the metal to superconductor transition in ultrasmall grains \cite{Duk00},
where the original Richardson's exact solution of the BCS model \cite{Richardson1963}
was rediscovered, to their generalization to a broad range of phenomena in interacting quantum
many-body systems \cite{Duk04, Ortiz2005}. The rational or XXX family of  integrable RG
models has been extensively studied, and includes the constant pairing Hamiltonian (BCS model)
\cite{Rich1966,Ortiz2005-2,Duk2006}, the central spin model \cite{Bortz2010}, generalized Tavis-Cummings
models \cite{Duke2004}, and more recently, open quantum systems \cite{Row2018}.   The
hyperbolic or XXZ family is much less investigated.  The notable $p + i p$ model of $p$-wave
fermionic pairing \cite{Sierra2009, Rombouts2010, VanNeck2014} is an exception, having
the Moore-Read (MR) Pfaffian, proposed for the non-Abelian quantum Hall fluid with
filling fraction 5/2  \cite{Moo91, Rea00}, as ground state at a given coupling strength.
Another recent finding is a number conserving version of the Kitaev wire
 which hosts topologically trivial and non-trivial superfluids phases \cite{Ortiz2014}.
Interestingly, its repulsive version in the strong coupling limit has been
shown to be related to the quantum Hall Hamiltonian projected onto the lowest Landau level subspace \cite{Ortiz2013}.

\begin{figure}[htb]
\includegraphics[width=.46\textwidth]{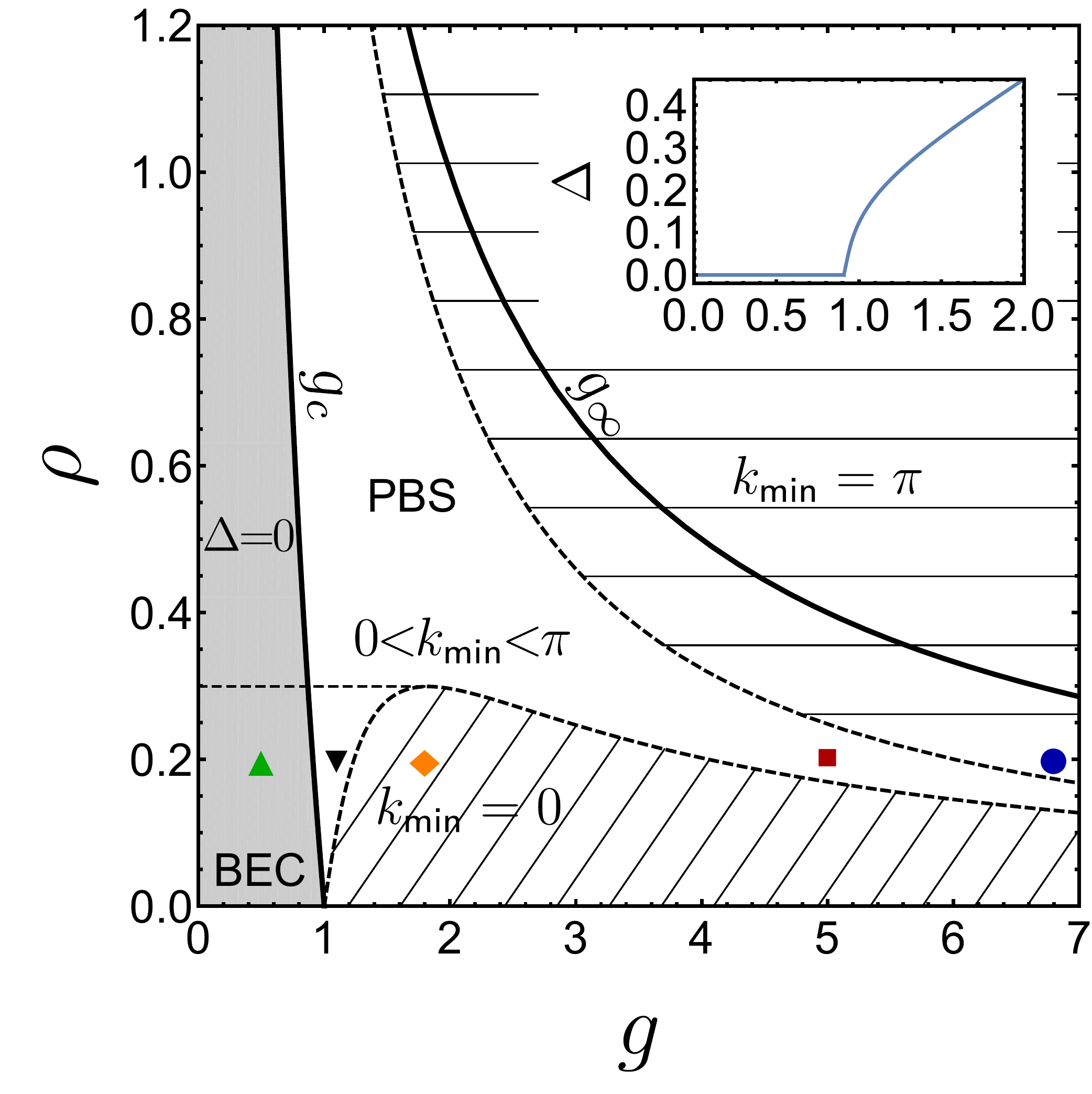}
\caption{Quantum phase diagram of the $p$-wave bosonic Hamiltonian, Eq. \eqref{Hphys} , 
in the space  $g=GL$ and $\rho=N/(2L)$, where $N$ is the number of bosons and $2L$ the size of the lattice. The gray area indicates the singlet
pair fragmented BEC phase, separated from the pair Bose superfluid phase (PBS) by a critical line $g_{c}$.
In the superfluid phase
the Volovik line, depicted by the lower dashed line, delineates the phase region where
the minimum of the quasi-particle energy is at momentum $k_{\sf min}=0$, while to the right of the upper dashed line
it is at $k_{\sf min}=\pi$.  Between these two lines  $0<k_{\sf min}<\pi$. All pairons diverge to infinity at $g_\infty$.
Symbols indicate the couplings used below at  density $\rho=0.2$. Horizontal dashed line is $\rho=0.299433$.
The inset shows the behavior of the pairing
gap $\Delta$ versus coupling strength $g$ at density $\rho=0.2$. }
\label{pd}
\end{figure}
Contrary to the fermionic case, $su$(1,1) bosonic RG models are unexplored territory.
Richardson introduced the bosonic constant pairing Hamiltonian \cite{Rich1967}, later generalized to study
condensate fragmentation for repulsive pairing interactions \cite{Schuck2001}, and the
transition from spherical to $\gamma$-unstable nuclei
in the nuclear interacting boson model \cite{Pan, Pittel}. The hyperbolic $su$(1,1)
RG, proposed in \cite{Dukelsky2001}, has only been employed to demonstrate
the integrability of the celebrated Lipkin-Meshkov-Glick
model in the Schwinger boson representation \cite{Ortiz2005, Pan2, Lerma2013}.  In this Letter we derive
an integrable two bosonic species $p$-wave pairing Hamiltonian, and study its quantum phase diagram.
We are motivated by the recent experimental observation of broad $p$-wave resonances in ultracold
$^{85}$Rb and $^{87}$Rb atomic mixtures \cite{Papp2008, Dong2016} that could lead to
stable thermodynamics phases dominated by $p$-wave attractive interactions. Mean-field
studies based on a two-channel model  predict three phases  \cite{Rad2009}: a) an atomic Bose-Einstein condensate (BEC)
for large negative detuning, b) a molecular BEC for large positive detuning, and c) an atomic-molecular BEC for
intermediate detuning. Quantum fluctuations may stabilize the
atomic-molecular phase for certain densities giving rise to the formation of polar superfluid droplets \cite{Li2019}.
Our exactly-solvable attractive one-channel $p$-wave model displays two phases (see Fig. \ref{pd}), a
gapless fragmented BEC of singlet pairs, where each of the species condenses into the lowest finite
momentum (grey area), and a gapped pair Bose superfluid (white area).
%A third-order quantum phase transition
%separates these two phases as in the integrable $p + i p$ one-channel \cite{Rombouts2010}
%and two-channel \cite{Lerma2011} fermionic models.
%At criticality, $g_{c}$,
%the exact ground state is a singlet pair BEC, a Bose analogue of the MR Pfaffian.
%\textcolor{red}{ We leave for a future work  a deeper study  of  the strong coupling limit,  $g>g_\infty$ (see definition below),    to determine whether  a crossover to a molecular BEC, similar to the $s$-wave fermionic BCS to BEC crossover  \cite{Ortiz2005-2},  is obtained in this bosonic case.}
%In the strong coupling limit, we expect a crossover to a molecular BEC similar to the $s$-wave fermionic BCS to BEC crossover \cite{Ortiz2005-2}.
%The dashed areas refer to different quasi-boson dispersions that we will discuss below.

\section{Hyperbolic $su$(1,1) RG integrals of motion}

The hyperbolic $su$(1,1) model for two bosonic species $a$ and $b$ in momentum ${\bf k}$ space is based
on the interspecies pair operators
\begin{eqnarray}
K_{\bold{k,Q}}^+\!\!&=&\!\! b_{\bold{k+Q}}^\dagger a_{-\bold{k}}^\dagger - a_{\bold{k+Q}}^\dagger b_{-\bold{k}}^\dagger,
\ \ \  K_{\bold{k,Q}}^-=(K_\bold{k,Q}^+)^\dagger , \label{su11}\\
 K_{\bold{k,Q}}^z\!\!&=&\!\! \frac{\hat{N}_{\bold{k,Q}}}{2}+1, \   \hat{N}_{\bold{k,Q}}=
 n^b_{\bf k +Q}+ n^b_{\bf -k}+ n^a_{\bf k +Q}+ n^a_{\bf -k} , \nonumber
 \end{eqnarray}
where $n^b_{\bf k}\! =b^\dagger_{\bold{k}}b^{\;}_{\bold{k}}$ and $n^a_{\bf k}\! =a^\dagger_{\bold{k}}a^{\;}_{\bold{k}}$.
 In order   to satisfy  the $su$(1,1) algebra $[K_{\bold{k,Q}}^-, K_{\bold{k',Q}}^+]=2\delta_{\bold{k},\bold{k'}}
 K_{\bold{k,Q}}^z$
and $[K_{\bold{k,Q}}^z, K_{\bold{k',Q}}^\pm]=\pm\delta_{\bold{k},\bold{k'}}K_{\bold{k,Q}}^\pm$, and to
avoid double counting,  we restrict  momenta  $\bf k$  and $\bf Q$ to have the component along  one of the
dimensions, for instance $k_x$, larger than zero,  $k_x>0$ and $Q_x>0$. As we will see below this does
not restrict the $\bf k$ values in the Brillouin zone.   The
operator $K_{\bold{k,Q}}^+$, that creates a two-species pair  with center-of-mass momentum $\bf Q$,
is antisymmetric under the exchange of species. If we interpret both species as the two
components of a pseudo-spin $1/2$, the pair operator $K_{\bold{k,Q}}^+$ creates a singlet state. The
pseudo-spin $1/2$ bosons define an independent and commuting $su$(2) spin algebra generated by
$S_{\bold{k,Q}}^z=(n^b_{\bf k +Q}+ n^b_{\bf -k}- n^a_{\bf k +Q}- n^a_{\bf -k})/2$,
 $S_{\bold{k,Q}}^+=b_{\bold{k+Q}}^\dagger a^{\;}_{\bold{k+Q}}
+b_{-\bold{k}}^\dagger a^{\;}_{-\bold{k}}$,  $S_{\bold{k,Q}}^-=(S_\bold{k,Q}^+)^\dagger$. Although we will
focus on the ${\bf Q}={\bf 0}$ case, these commuting algebras can be exploited to describe
Larkin-Ovchinnikov-Fulde-Ferrell-type phases and/or mass imbalance two-component cold atom gases as
described in \cite{Duk2006} for fermionic systems.

In terms of the $su$(1,1) generators (\ref{su11}), the  hyperbolic integrals of motion for ${\bf Q}={\bf 0}$
are  \cite{Dukelsky2001, Ortiz2005}
\begin{eqnarray}
R_\bold{k}&=&K_{\bold{k}}^z-2 \lambda \sum_{\bold{k}'(\not=\bold{k})>0}\left[
\frac{\eta_\bold{k}\eta_{\bold{k}'}}{\eta_\bold{k}^2-\eta_\bold{k'}^2}\left(K_{\bold{k}}^+ K_{\bold{k}'}^-+
K_{\bold{k}}^-K_{\bold{k}'}^+\right)
\right. \nonumber\\
& &\left.
-\frac{\eta_\bold{k}^2+\eta_{\bold{k}'}^2}{\eta_\bold{k}^2-\eta_{\bold{k}'}^2} K_\bold{k}^z K_{\bold{k}'}^z\right],
\label{Integral}
\end{eqnarray}
where $\eta_\bold{k}$ are arbitrary odd functions of ${\bf k}$. The sum $\bold{k}'>0$ means
that the component $k'_x$ should be positive.

For a fixed number of bosons $N=2M+\nu$, where $M$ is the number of singlet boson pairs and $\nu$
the total number of unpaired bosons, the  eigenvalues of the integrals of motion are
\begin{eqnarray*}
r_\bold{k}=d_{\bold{k}}\left[1+2 \lambda \sum_{\bold{k}'(\not= \bold{k})>0}d_{\bold{k}'}\frac{\eta_{\bold{k}}^2+
\eta_{\bold{k}'}^2}{\eta_{\bold{k}}^2-\eta_{\bold{k}'}^2}-2 \lambda \sum_{\alpha=1}^M\frac{e_\alpha+
\eta_{\bold{k}}^2}{e_\alpha-\eta_{\bold{k}}^2}\right],
\end{eqnarray*}
where $d_\bold{k}=\nu_\bold{k}/2+1$, $\nu_{\bold{k}}$  is the seniority quantum number (number of
unpaired bosons) of level $\bold{k}$, and $\nu=\sum_{\bold{k}>0} \nu_{\bold{k}}$. The spectral
parameters $e_\alpha$, so-called pairons, are roots of the Richardson equations
$(\alpha=1,\dots,M)$
\begin{equation}
\sum_{\bold{k}>0}
\frac{d_{\bold{k}}}{\eta_{\bold{k}}^2-e_\alpha}+\sum_{\beta=1 (\beta\not=\alpha)}^M \frac{1}{e_{\beta}-
e_\alpha}+ \frac{\tilde Q}{e_\alpha}=0 ,
\label{RichEq}
\end{equation}
with
\begin{eqnarray*}
\tilde Q=-\frac{1}{4 \lambda}+\frac{M-1+\sum_{\bold{k}>0} d_{\bold{k}}}{2}.
\end{eqnarray*}

Each independent solution of the Richardson equations (\ref{RichEq}) defines a common  eigenstate
of the integrals of motion (\ref{Integral}):
\begin{equation}
    |\Phi_{M,\nu}\rangle=\prod_{\alpha=1}^M \left( \sum_{\bold{k}>0} \frac{\eta_{\bold{k}}}{\eta_{\bold{k}}^2-
    e_\alpha} K_{\bold{k}}^+ \right)|\nu\rangle ,
\label{wave}
\end{equation}
where the state $|\nu\rangle$, with $\nu$ unpaired bosons, satisfies $\hat{K}_\bold{k}^-|\nu\rangle=0$
for all $\bold{k}$, and  $\hat{K}_\bold{k}^z|\nu\rangle=d_\bold{k}|\nu\rangle$.

By combining the integrals of motion $R_{\bf k}$ with the Hellmann-Feynman theorem \cite{Rombouts2010},
the occupation probabilities can be obtained from the expectation value
%of $K_{\bold k}^z$
\begin{equation} \hspace*{-0.2cm}
\left\langle \Phi_{M,\nu} | K_{\mathbf{k}}^{z}| \Phi_{M,\nu}\right\rangle =d_{\mathbf{k}}\left(
1-2\lambda^{2}\sum_{\alpha=1}^{M}\frac{2\eta_{\mathbf{k}}^{2}}{\left(
\eta_{\mathbf{k}}^{2}-e_{\alpha}\right)  }\frac{\partial e_{\alpha}}
{\partial\lambda}\right) ,
\label{occuprob}
\end{equation}
where the pairon derivatives can be obtained from the derivatives of Eq. (\ref{RichEq}) leading to a linear set of equations.
For ease of presentation we consider next a one-dimensional version of the $p$-wave model.
It is straightforward to extend our model to higher dimensions as has been done in the
fermionic case \cite{Sierra2009, Rombouts2010}.

\section{The $p$-wave Bose Hamiltonian}

The $p$-wave pairing Bose Hamiltonian we want to study is given by
\begin{eqnarray}   \label{Hphys}
H&=&\sum_k \eta_k^2\left( a^\dagger_k a^{\;}_k+b_k^\dagger b^{\;}_k\right) \\
  & &\hspace*{-0.5cm} -\frac{G}{4}\sum_{k, k'} \eta_k\eta_{k'} \left(b_k^\dagger a_{-k}^\dagger-a_k^\dagger
  b_{-k}^\dagger\right)\text{\Large$($} b^{\;}_{k'} a^{\;}_{-k'}-a^{\;}_{k'} b^{\;}_{-k'}\text{\Large$)$}\nonumber
\end{eqnarray}
%\begin{equation}
%H=\sum_{k>0}\eta_k^2 \, \hat{N}_{k}-G\sum_{k,k^{\prime}>0}\eta_k \eta_{k'} K_{k}^{+}K_{k^{\prime}}^{-} ,
%\label{H}
%\end{equation}
where $\eta_k=\sin(k/2)$ and $\eta_k^2=(1-\cos k)/2$. Assuming antiperiodic boundary
conditions the allowed $k$ values are $k= \pm \pi/2 L, \pm 3\pi/2 L, \dots , \pm(2 \pi L-\pi)/2 L $,
with $2 L$ the size of the chain and $L$ the number of $su$(1,1) copies.  We have chosen antiperiodic
boundary conditions to explicitly exclude the $k=0$ state. This state cannot support singlet pairs and,
therefore, it will be excluded from the dynamics of $p$-wave pair scattering.
This model Hamiltonian, which  written in terms of the  $su(1,1)$ generators is
\begin{equation}
H=\sum_{k>0}\eta_k^2 \, \hat{N}_{k}-G\sum_{k,k^{\prime}>0}\eta_k \eta_{k'} K_{k}^{+}K_{k^{\prime}}^{-}, \label{H}\\
\end{equation}
can be derived from the hyperbolic $su$(1,1) RG  integrals of motion
(\ref{Integral}), by using the linear combination
%$$
%H=8 x \sum_{k>0} \eta_{k}^2 R_{k}-8Z-4G\sum_{k>0} d_k (1-d_k) \eta_{k}^2 - 2\sum_{k>0} \hat{N}_k,
% $$
\begin{equation}
H=2 x \sum_{k>0} \eta_{k}^2 R_{k}-2Z-G\sum_{k>0} d_k (1-d_k) \eta_{k}^2,\nonumber
\end{equation}
where $x=[1+2\lambda(M+L+(\nu/2)-1)]^{-1}$, $Z=\sum_{k>0} \eta_k^2$, and  $G= 4
\lambda/(1+2 \lambda (M-1+\sum_{k>0} d_{k}))$.

Our $p$-wave Hamiltonian \eqref{H} has an explicit U(1) symmetry, i.e., conservation of the total
number of bosons, and
a pseudospin invariance that basically preserves the polarization $S^z=\sum_k S^z_k$, that is,
the difference between the number
of bosonic species. Here we will focus on an unpolarized mixture of atoms characterized by $S^z=0$, although the
polarized case ($S^z\neq 0$) is contained in our exact solution. For instance, an excess of $a$ atoms manifests
through the seniorities $\nu_k$ specifying the  $k$ states occupied by the unpaired $a$ atoms.

Eigenvalues of (\ref{H}) can be determined from the
 integrals of motion, using the same linear combination, which, after using Eq. (\ref{RichEq}), gives
\begin{equation}
%E=\sum_{k>0}\varepsilon_{k} \nu_{k}+8\sum_{\alpha=1}^M e_\alpha- 4 M.
E=\sum_{k>0}\eta_{k}^2 \nu_{k}+2\sum_{\alpha=1}^M e_\alpha.
\label{Energy}
\end{equation}

\begin{figure}
\includegraphics[width=.42\textwidth]{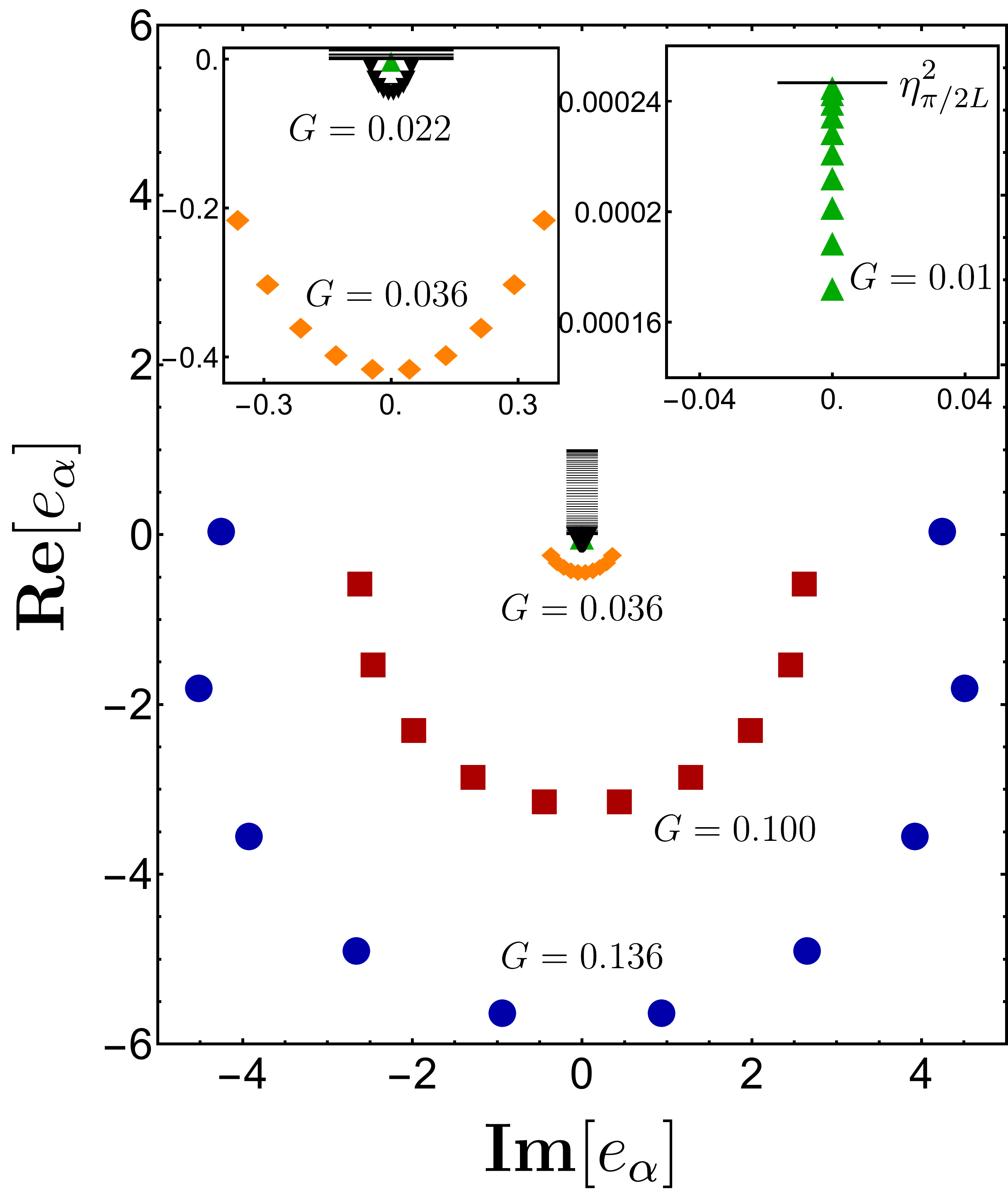}
\caption{Pairons, $e_\alpha$,   for a finite system with $M=10$ pairs and $L=50$ levels
($\rho=M/L=0.2$). Cases displayed correspond to the five symbols indicated in Fig. {\ref{pd}}.
Coupling strengths $G $ are indicated by numbers close to the respective symbols.
% $G=0.01$ (upside triangles), $G=0.022$ (downside triangles), $G=0.036$ (diamonds), $G=0.1$
%(squares), and $G=0.136$ (circles).
Horizontal lines depict  $\eta_k^2$ levels. See \cite{SM} for an animation of the pairons evolution as a function
of $G$ for $0<G<G_\infty$.}
\label{pairons}
\end{figure}

Let us analyze next the way the
ground state evolves as a function of coupling strength $G\ge0$ (Fig. \ref{pairons}).
Each independent solution of the Richardson equations (\ref{RichEq}) provides a set of $M$
pairons that define both the energy eigenvalue (\ref{Energy}) and the corresponding eigenstate
(\ref{wave}). The ground state (with $\nu=0$) for weak coupling $G$ has the pairons distributed in the
real interval between zero and the minimum $\eta_{\pi/2L}^2=\sin^2(\pi/4 L)$.
At $G_{n}=2/(2L+2M-n-1)$,
$n$ pairons collapse to zero. In between collapses, the $n$ pairons
expand as complex conjugate pairs forming an arc in the complex plane
around zero. The whole set of  $M$ pairons collapses to zero at the critical point
%(as we will see below)
\begin{eqnarray*}
G_{c}=\frac{2}{2L+M-1},
\end{eqnarray*}
where the exact (non-normalized) ground state becomes a condensate of singlet pairs
\begin{equation}
\left\vert \Phi_{M}\right\rangle _{\sf BMR}=\left(  \sum_{k>0}\frac{1}{\eta_{k}
}K_{k}^{+}\right)^{M}\left\vert 0\right\rangle ,
\label{MR}
\end{equation}
which  is algebraically analogous to the MR state of the $p+ip$ fermionic model \cite{Sierra2009, Rombouts2010},
and, therefore, we will call it Bose Moore-Read (BMR) state.

Naively, in an extended system, one would expect that the ground state of the BEC
phase, $0\le G \le G_c$,  corresponds to a zero-momentum condensate for each species
\begin{equation}
\left\vert \Phi\right\rangle =a_{0}^{\dagger M}b_{0}^{\dagger M}\left\vert
0\right\rangle
\label{Ferro}
\end{equation}
since, as we will see, the quasi-particle gap $\Delta$ vanishes. This state has maximum spin $S=M$.
For mesoscopic systems, it has been shown that the correct ground state at weak coupling is a fragmented singlet pair
BEC  \cite{Kuklow, Leggett}, which in  momentum space becomes
\begin{equation}\hspace*{-0.25cm}
\left\vert \Phi\right\rangle =(K^+_{k_{\sf min}})^M \!\! \left\vert 0\right\rangle =\left(  b_{k_{\sf min}}^{\dagger}a_{-k_{\sf min}
}^{\dagger}- a_{k_{\sf min}}^{\dagger}b_{-k_{\sf min}}^{\dagger}\right)
^{M}\! \! \left\vert 0\right\rangle ,
\label{singlet}
\end{equation}
with $k_{\sf min}=\pi/2L$ for the antiperiodic chain. Note that in this phase, the exact ground state has  a mixture
of complex pairons close to zero and real pairons in the  interval $[0,\eta^2_{\pi/2L}]$. %$0<e_\alpha <  \eta^2_{\pi/2L}$.
For large  $L$ the pairons will cluster around zero and the exact ground state (\ref{wave}) will tend to the BMR state
(\ref{MR}) which is representative of the whole phase. The BMR state is controlled by $k_{\sf min}$, and
therefore it converges to the singlet pair condensate in the large $L$ limit.
Interestingly, in the thermodynamic limit the states (\ref{singlet}) and (\ref{Ferro}), as well as condensates
with other spin quantum numbers $S$, become degenerate. A weak repulsive
interaction may destabilize those degenerate spin states against the singlet pair condensate \cite{Leggett}.

For $G>G_{c}$ the pairons distribute along an arc that expands in the complex plane as  $G$ increases
(Fig. \ref{pairons}). At
\begin{eqnarray*}
G_\infty=\frac{2}{M-1}
\end{eqnarray*}
the absolute value of all pairons diverges to infinity. This divergence does not affect the energy since
imaginary parts cancel out pairwise in ($\ref{Energy}$) and the real parts combine to give
$E=2G M \sum_{k>0} \eta^{2}_{k}$. Infinite pairon energies have been observed previously in fermionic
hyperbolic models \cite{Ortiz2014} and they were related to a duality associated to the
particle-hole symmetry \cite{Links}. At this point the exact ground
state can be expressed as a different pair condensate
\begin{equation}
    |\Phi_{M}\rangle_{ G_\infty}=\left( \sum_{k} \eta_{k}\, K_{k}^+ \right)^M|0\rangle .
\end{equation}
In turn, in the fermionic case we find that this state appears as the highest energy eigenstate in the repulsive pairing region.

In Fig. {\ref{pd}},   at density $\rho=0.2$,  we show five distinct symbols covering all distinct areas of the phase diagram,
at couplings $g=0.5, 1.2, 1.8, 5.0, 6.8$, with $g=G L$. Figure {\ref{pairons}} displays pairons of a finite-sized system
with $M=10$ and $L=50$,  for these
same five values. % ($G=g/L$)
As discussed above, the first point with $G<G_{c}$ has 10 pairons distributed in the
real positive axis below $\eta^2_{\pi/2 L}$ (see the right inset). After the pairons collapse to zero at
$G_{c}$, they form an arc in the complex plane that expands for increasing values of $G$.
This is the case for the remaining four couplings that lay in between $G_{c}$ and $G_{\infty}$,
two of them can be seen in the left inset while the other two in the central figure.

\begin{figure}[h]
\includegraphics[width=.46\textwidth]{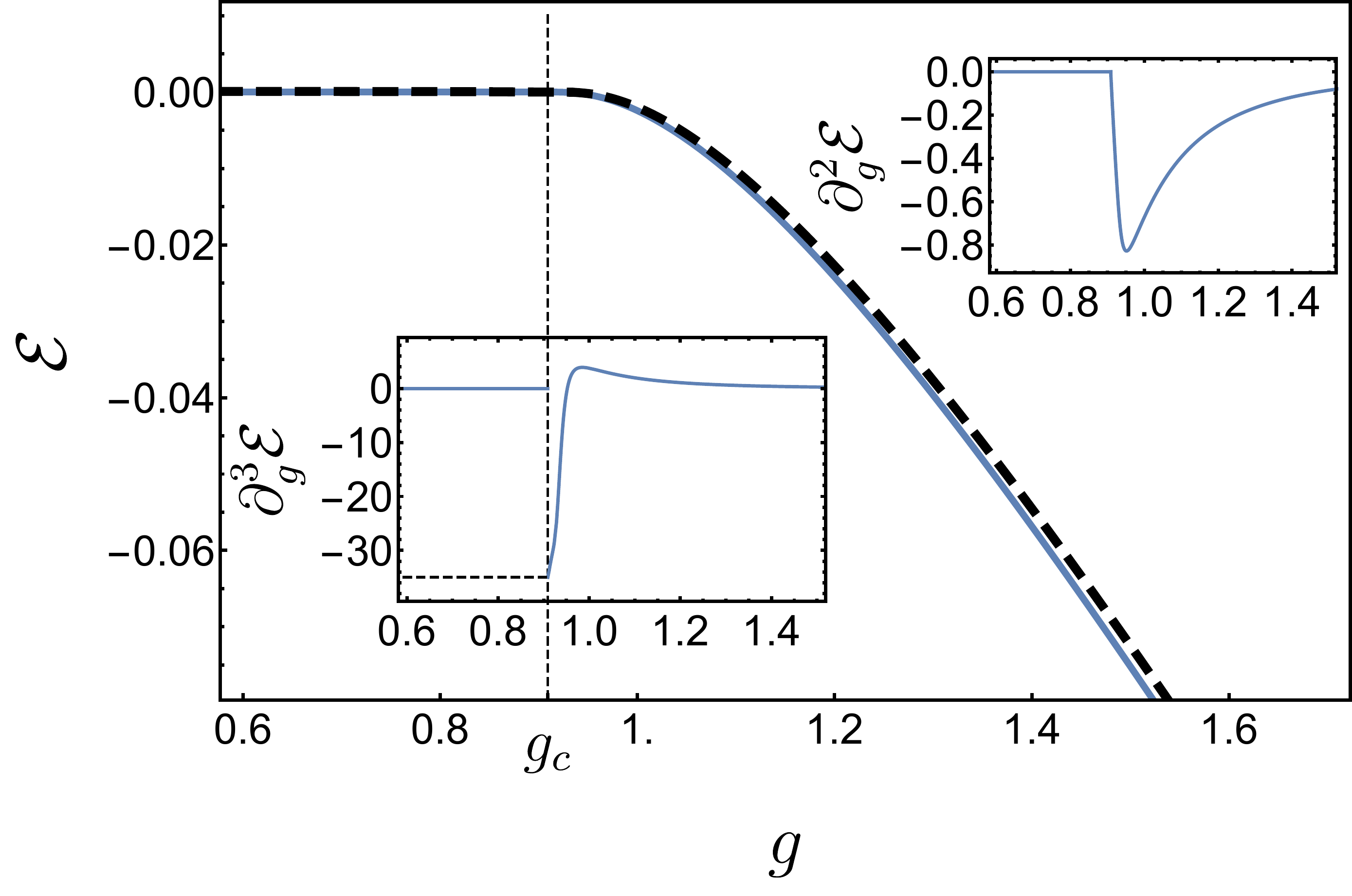}
\caption{Energy density ${\cal E}$ as a function of $g$. Continuous lines are the thermodynamic limit solution
while the dashed line is the exact ${\cal E}$ for $M=10$ and $L=50$. Second and third-order derivatives of $\cal E$ are
displayed in the insets.}
\label{energydensity}
\end{figure}

\section{Quantum Phase Diagram}

The thermodynamic limit is obtained in the limit of $N,L\rightarrow \infty$
with constant density $\rho=N/(2L)$ and rescaled interaction
strength $g=G L$. In this limit, the Richardson equations (\ref{RichEq}) transform into the boson
gap and number equations \cite{Rombouts2010, Ortiz2005-2}
\begin{align}
\frac{\pi}{g}=\int_{0}^{\pi}\frac{\eta_{k}^{2}}{E_{k}} \, dk,\;\;\ \ \ \
\rho=\frac{2}{\pi}\int_{0}^{\pi}v_{k}^{2} \, dk,
\label{gapnumber}
\end{align}
with quasi-boson energies $E_k$ and occupation probabilities $v_k^2$
\begin{align}\hspace*{-0.2cm}
E_k=\sqrt{\left({\eta_k^2}-\mu \right)^2-4\eta^2_k \Delta^2},\;\;
v^2_k=\frac{1}{2} \left (\frac{\eta_k^2-\mu}{E_k}-1\right ),
\label{quasi}
\end{align}
where $\mu$ is the chemical potential and $\Delta$ the gap. Though  $E_k$ in (\ref{quasi}) may, in principle,
be complex, we have numerically verified that in the large attractive $g$ limit, Eqs. (\ref{gapnumber})
have solutions $\mu\approx -\gamma_1 g$ and $\Delta \approx \gamma_2 g$, with  $\gamma_{1,2}$ positive constants satisfying
$4 \gamma_2^2<\gamma_1^2$. This latter condition  guarantees that the quasi-boson energies, given by
$E_k\approx g \gamma_1\sqrt{1-(4\gamma_2^2/\gamma_1^2)\eta_k^2}$ are always real,  even in the limit $g\rightarrow\infty$.
The ground state energy density ${\cal E}\equiv E/L$ for a given density $\rho$ in the thermodynamic limit is given by
\begin{eqnarray}
%\frac{{\cal E}}{4}+\rho
{\cal E}=-\frac{4 \Delta^2}{g}-1 +\frac{2}{\pi} \int_{0}^{\pi}\frac{\eta_{k}^{2}(\eta_{k}^{2}-\mu)}{E_{k}} \, dk .
\label{Energydensity}
\end{eqnarray}

The critical coupling of the exact solution in the finite-size case, becomes
$g_{c}= \lim_{L,N\rightarrow\infty} [G_{c} L]=2/(2+\rho)$ in the thermodynamic limit.
%This critical coupling,
%indicated by a solid line in Fig. {\ref{pd}}, marks the transition to a pair Bose superfluid phase.
The gap
$\Delta$ is zero at weak pairing up to the critical value $g_{c}$. The inset of Fig. {\ref{pd}} shows
the behavior of the gap for $\rho=0.2$. It increases monotonically for $g>g_{c}$.
In the same thermodynamic limit,  the coupling where all pairons diverge becomes $g_{\infty}=
\lim_{L,N\rightarrow\infty} [G_{\infty} L]=2/\rho$ (Fig. {\ref{pd}}).
%, also indicated in Fig. {\ref{pd}} by a solid line.

We are interested in establishing the nature of the non-analyticities of ${\cal E}$ at the critical point.
It turns out that ${\cal E}=0$,
for $0< g<g_c$ and is non-analytic at $g=g_c$ with  a third order phase transition to a pair superfluid phase
\cite{ Rombouts2010,Lerma2011}.
Close to $g-g_c\approx 0^+$, it behaves as
\begin{eqnarray}
{\cal E} \approx -\frac{\pi^2}{3}  \left(\frac{\tilde{g}}{g} \right)^2 \left (\frac{\tilde{g}}{g} - 12 \frac{g-1}{g e^{2(2-g_c)}}
\ e^{\frac{2 (g_c-1)}{\tilde{g}} }\right) ,
\label{NAEnergydensity}
\end{eqnarray}
where $\tilde{g}=(g-g_c)/g_c$ (See Appendix \ref{AppendixA}). Interestingly, the behavior of $\cal E$ close to $g_c$
depends on $\rho$ only through its critical value $g_c$.
The first and second-order derivatives at the critical point are zero, while the third-order derivative is
$\partial_g^3 {\cal E} \rfloor_{g-g_c\rightarrow 0^+} =-2 \pi^2/g_c^6$, signaling a discontinuity of third order.
This is illustrated in Fig. \ref{energydensity} for $\rho=0.2$ where, moreover, ${\cal E}$  is compared with the
exact energy density for $M=10$ and $L=50$.
%The insets display the second and third-order derivatives of the
%energy confirming the discontinuity in the third-order derivative.

\begin{figure}[h]
\includegraphics[width=.46\textwidth]{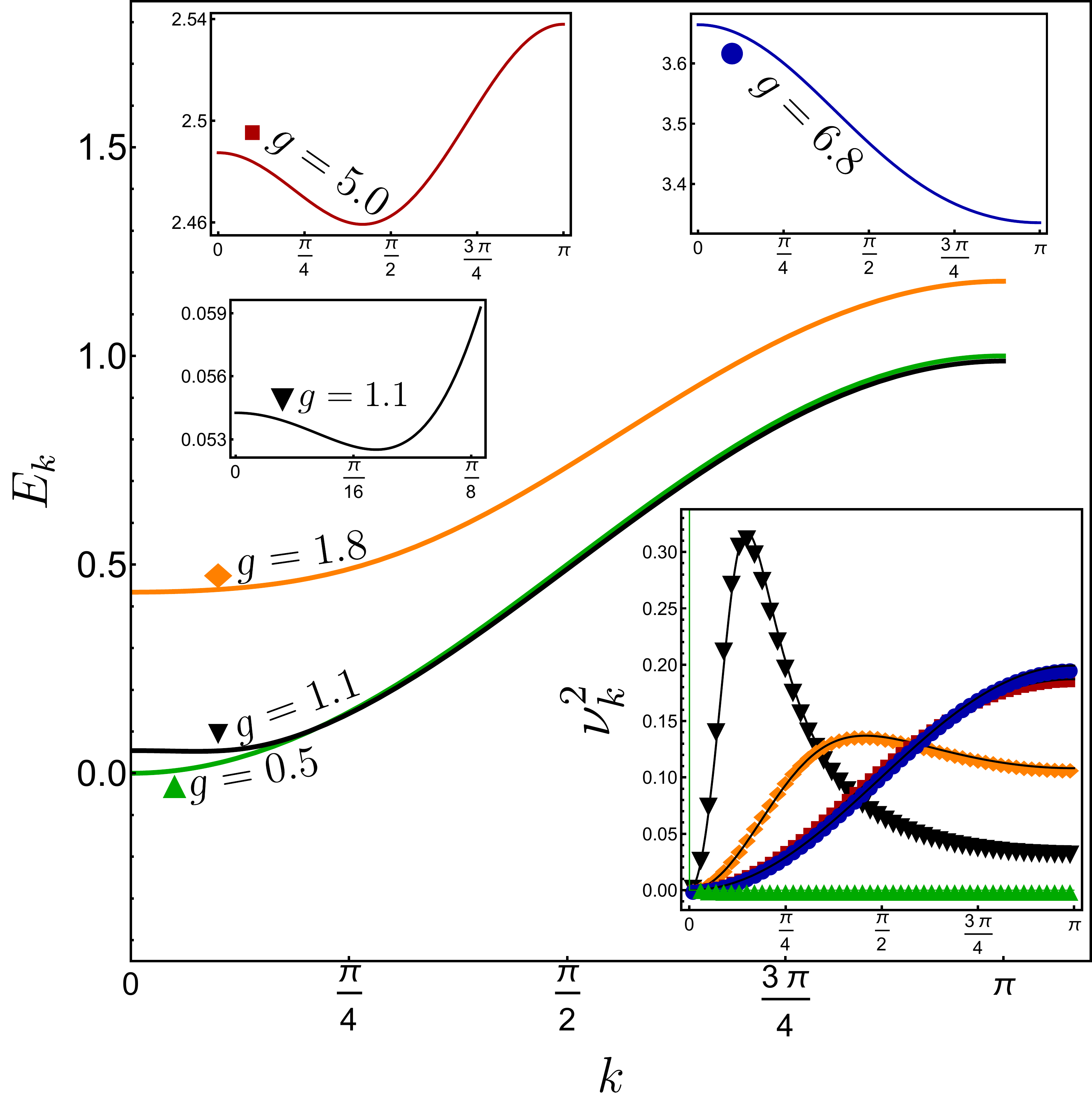}
\caption{Quasi-particle energies for $\rho=0.2$ and the same couplings as those indicated in  Fig. \ref{pd}.
For $g=1.1$, the inset zooms in the low-$k$ region, showing that $k_{\sf min}\not= 0$. Lower inset
displays the occupation probabilities. }
%\caption{Quasi-particle energies for $\rho=0.2$ and the same couplings as those indicated in  Fig. \ref{pd}.
%Lower inset shows a zoom to the low-$k$ region of the coupling $g=1.1$, showing that for this case $k_{\sf min}\not= 0$. }
\label{quasiboson}
\end{figure}

\section{Nature of excitations}

In Fig. \ref{quasiboson} we show the quasi-boson energies for the  five values of $g$ indicated in
Fig. \ref{pd}.  The quasi-boson energies
change from  $E_k=\sin^2(k/2)$ in the gapless pair condensate phase ($g=0.5$), to a complex dispersion
in the pair Bose superfluid phase. For $\mu+2\Delta^2\leq 0$,  $E_k$ is a monotonous increasing function
with minimum at  $k_{\sf min}=0$ and energy $E_{k_{\sf min}}=|\mu|$ ($g=1.8$). The previous condition
is fulfilled in the superfluid phase only for small densities $\rho<0.299433$ in a finite coupling interval.
The region  is indicated  by the area with diagonal lines in Fig. \ref{pd}. The boundary of this region,
the so called Volovik line \cite{Sierra2009} defined by a superfluid with the minimum quasi-boson energy
at $k=0$, is given by $\mu+2\Delta^2=0$. For $0<\mu+2\Delta^2<1$, $E_k$ has a minimum  at
$k_{\sf min}=2\arcsin(\sqrt{\mu+2\Delta^2})$, satisfying  $0<k_{\sf min}<\pi$ ($g=1.1, 5.0$). The region
of the phase diagram where $E_k$ has this dispersion is indicated by the white area in Fig. \ref{pd}.
The previous condition is fulfilled for any density, and gives  the form of the quasi-boson dispersion
immediately after the quantum phase transition. For $\mu+2\Delta^2\geq 1$ (area with horizontal lines
in Fig. \ref{pd}), the quasi-boson dispersion is a monotonous decreasing function with minimum at
$k_{\sf min}=\pi$ ($g=6.8$).

The occupation probabilities in momentum space are displayed in the lower inset of
Fig. \ref{quasiboson} for the five
values of $g$ indicated in Fig. \ref{pd}. Continuous lines are the thermodynamic limit solution
and symbols correspond to the exact
solution for the finite-size case calculated using Eq. \eqref{occuprob}.
For $g=0.5$ the system is condensed in $k_{\sf min}$ resulting
in a delta distribution in the thermodynamic limit. At  $g_c$, in that limit,
the macroscopic occupation at $k_{\sf min}\rightarrow 0$ jumps to zero and then the maximum of the
distribution moves to finite $k$ values. This jump in the $k=0$ momentum state resembles the
one observed in the $p+ip$ and RG Kitaev models \cite{Rombouts2010,Ortiz2014} and the $s$-$d$ RG model of
Ref. \cite{Claeys2018}. In the fermionic case, 
this fact has been linked to a topological phase transition  \cite{Rombouts2010,Ortiz2014}.
For $g=1.1$ and $1.8$  the profiles broaden
and maxima get displaced to larger values of $k$. Finally, for $g=5.0$ and $6.8$ the profiles
are inverted with a maximum occupation at $k=\pi$.

%\begin{figure}[h]
% \includegraphics[width=.42\textwidth]{fig5.pdf}
%\caption{Occupation probabilities in the thermodynamic limit (continuous lines) and from the exact
%Richardson solution (symbols) for the five values of $g$ highlighted in Fig. \ref{pd}.}
%\label{occupation}
%\end{figure}

\section{Outlook}

We introduced an exactly-solvable two species $p$-wave
bosonic model and established its quantum phase diagram in the attractive sector.
Only the case of a balanced mixture with equal masses ($m_a=m_b$) and zero center-of-mass
momentum ${\bf Q}$ has been studied in depth. Imbalanced binary mixtures ($\nu\neq 0$,
$m_a \ne m_b$) and finite ${\bf Q}$ pairs are contained within our exactly solvable model.
The exact, finite and thermodynamic limit, treatments of the $p$-wave pairing
Bose Hamiltonian (\ref{H}), although seemingly similar, have profound physical
differences  to its fermionic
counterpart \cite{Sierra2009, Rombouts2010, VanNeck2014, Links}
despite the fact that both cases share a third-order quantum phase transition.
In the fermionic case the latter separates two gapped superfluid phases and has a topological
character \cite{Ortiz2014}. In the bosonic case one of the phases is gapless and displays a fragmented
BEC condensate with macroscopic occupations of both species in the lowest finite momentum
pair states $(-k, k)$, while the other is a gapped pair Bose superfluid (PBS). Moreover, while for fermions
the critical coupling takes place at the Read-Green point, with one pairon at zero energy and the
other $M-1$ pairons with real and negative energies, for bosons the phase transition takes
place at the equivalent of the fermionic Moore-Read point with all pairons collapsing to zero energy.
It is at this critical point that the exact bosonic ground state is a pair condensate
with amplitudes fixed by the single particle energies.
%Classes of states emerging from the model include a fragmented singlet pair  BEC, a
%Bose Moore-Read analogue and   a new pair Bose superfluid (PBS).
%{\bf and, for $g>g_\infty$, a molecular BEC with all real and negative pairons indicating bound pairs of the two species}.

 Motivated by a theoretical
prediction \cite{Petrov},  recent experiments unveiled a
new type of ultradilute quantum liquid in ultracold bosonic systems. Apparently, there is no unique mechanism
leading to such a liquid state since it has been observed in single-species dipolar systems \cite{Kadau} and Bose
(potassium) mixtures \cite{Semeghini,Cabrera}. Can one obtain a quantum liquid phase in $p$-wave Bose systems?
This question has been recently addressed in \cite{Li2019}, and answered in the affirmative for a particular model.
Our PBS represent a (fixed-point) number-conserving candidate for such quantum liquid phase.
The pairing interaction in \eqref{H} may thus provide a new
effective mechanism for its emergence. Although the superfluid gap protects that state from expansion in finite geometries,
further studies in trapped potentials are required to identify
a possible self-bound quantum liquid droplet. On the experimental side, it is crucial to have a precise understanding of
 the spectrum of excitations to  compare to our theoretical predictions.

{\it Acknowledgments---}
S.L.-H. acknowledges financial support from the Mexican CONACyT project CB2015-01/255702.
J.D. is supported by the Spanish Ministerio de Ciencia, Innovaci\'on y Universidades,
and the European regional development fund (FEDER) under Projects No. FIS2015-63770-P and
PGC2018-094180-B-I00, S.L.-H. and J.D. acknowledges financial support .from the Spanish collaboration Grant I-COOP2017 Ref:COOPB20289. G.O acknowledges
support from the US Department of Energy grant  DE-SC0020343.

%J.D. acknowledges financial support from the Spanish Ministerio de Economía y Competitividad and the European
%regional development fund (FEDER) under Projects No. FIS2015-63770-P and PGC2018-094180-B-I00. The collaboration
%was supported by the Spanish Grant I-COOP2017 Ref:COOPB20289.

\appendix

\section{Non-analytic behavior at the  quantum critical point}
\label{AppendixA}

One can write the boson gap and number equations \eqref{gapnumber}, in the
thermodynamic limit,  as
\begin{eqnarray}
\pi (\rho +1) &=&  \int_0^1 d x \frac{x-\mu}{\sqrt{x-x^2}\sqrt{x^2-2 a x +\mu^2}} \\
\frac{\pi}{g} &=& \int_0^1 d x \frac{x}{\sqrt{x-x^2}\sqrt{x^2-2 a x +\mu^2}} ,
\end{eqnarray}
where the following change of variables has been performed: $x=\eta_k^2$, and
$a=\mu+2\Delta^2$.

We are interested in characterizing the behavior of physical quantities, such as the chemical potential $\mu$,
superfluid gap $\Delta$, and ground state energy density $\cal E$,  near the
phase transition $g\approx g_c$ where a non-analyticity develops. Close to the transition, and for
couplings $g>g_c$,  $\mu<0$ and
$a >0$, such that $a> \delta=\mu^2-a^2>0$. We need to determine the behavior of the
above integrals  in the limit $\delta \rightarrow 0^+$. A few algebraic steps lead to:
\begin{eqnarray}
\hspace*{-0.6cm}\int_0^1 d x \frac{1}{\sqrt{x-x^2}\sqrt{x^2-2 a x +\mu^2}} \stackrel{\delta \rightarrow 0^+}{=}  I_d \nonumber\\
I_d  =\sqrt{\frac{1}{a-a^2}} \left ( \log 16 + 2 \log (a-a^2) - \log \delta \right ) .
\end{eqnarray}
Similar manipulations result in
\begin{eqnarray}
\int_0^1 d x \frac{x-a}{\sqrt{x-x^2}\sqrt{x^2-2 a x +\mu^2}}  &\stackrel{\delta \rightarrow 0^+}{=}& \pi - 4 \arcsin \sqrt{a}.
\nonumber
\end{eqnarray}

Therefore, the resulting gap and number equations close to the critical point become
\begin{eqnarray}
\pi \rho  &=& - 4 \arcsin \sqrt{a} + (a-\mu)\,  I_d \\
\frac{\pi}{g} &=& \pi - 4 \arcsin \sqrt{a} + a \,  I_d,
\end{eqnarray}
or equivalently
\begin{eqnarray}
\rho  + 1 &=& \frac{1}{g}- \frac{\mu}{a} \left (\frac{1}{g} -1 +\frac{4}{\pi} \arcsin \sqrt{a}\right ) ,
\end{eqnarray}
and whose consistency can be checked by taking the limit $a\rightarrow 0$, $g \rightarrow g_c$. This gives
$\rho+1=2/g_c$, as expected from the exact solution. On the other hand, we would like to
determine the behavior of the gap $\Delta$ and chemical potential $\mu$ as a function of
$\rho$ and $g$ close to the transition. It turns out to be more convenient to write
$a-\Delta^2=\mu+\Delta^2=  \mu \, \Gamma$, and find solutions for $\mu$ and $\Gamma$
\begin{eqnarray}
\mu  &\approx& - \left (\frac{\pi}{2} \right )^2 \left (\frac{g-g_c}{g \, g_c} \right )^2\\
\Gamma &\approx&  4 \, e^{2(g_c-2)} e^{\frac{2g_c(g_c-1)}{g-g_c}},
\end{eqnarray}

What is the behavior of the ground state energy density $\cal E$, Eq. \eqref{Energydensity},
\begin{eqnarray}
{\cal E} &=& -\frac{4 \Delta^2}{g}-1+ \frac{2}{\pi} \int_0^1 d x \frac{x(x-\mu)}{\sqrt{x-x^2}\sqrt{x^2-2 a x +\mu^2}}
\nonumber
\end{eqnarray}
close to the phase transition? Following the same strategy, close to the transition point,
%\begin{eqnarray}
% \int_0^1 \!\! d x \frac{x(x-\mu)}{\sqrt{x-x^2}\sqrt{x^2-2 a x +\mu^2}}\stackrel{\delta \rightarrow 0^+}{=}
%  \frac{\pi}{2} \left [1+2(a(\rho+1) \right . \nonumber \\
%\left . -\mu) \right ] +2 \left ( \sqrt{a(1-a)} + (2\mu-1) \arcsin \sqrt{a} \right ) ,
%\nonumber
%\end{eqnarray}
\begin{eqnarray}
 \int_0^1 \!\! d x \frac{x(x-\mu)}{\sqrt{x-x^2}\sqrt{x^2-2 a x +\mu^2}}\stackrel{\delta \rightarrow 0^+}{=}
  \frac{\pi}{2} \left [2(a(\rho+1)-\mu) \right . \nonumber \\
\left . + 1 \right ] +2 \left ( \sqrt{a(1-a)} + (2\mu-1) \arcsin \sqrt{a} \right ) ,
\nonumber
\end{eqnarray}
and to first order in powers of $\Gamma$, it results
\begin{eqnarray}
{\cal E} \approx -\frac{\pi^2}{3}  \left(\frac{\tilde{g}}{g} \right)^2 \left (\frac{\tilde{g}}{g} - 12 \frac{g-1}{g e^{2(2-g_c)}}
\ e^{\frac{2 (g_c-1)}{\tilde{g}} }\right) ,
\end{eqnarray}
where $\tilde{g}=(g-g_c)/g_c$, displaying a discontinuity of third order as indicated in the main text.

\end{document}